# Deep learning with reflection high-energy electron diffraction images to predict cation ratio in $Sr_{2x}Ti_{2(1-x)}O_3$ thin films

*Sumner B. Harris[1]*[†], Patrick T. Gemperline[2][†], Christopher M. Rouleau[1], Rama K. Vasudevan[1], Ryan B. Comes[2,3]**

1. Center for Nanophase Materials Sciences, Oak Ridge National Laboratory, Oak Ridge, Tennessee 37831, United States.
2. Department of Physics, Auburn University, 315 Roosevelt Concourse, Auburn, Alabama 36849, United States.
3. Department of Materials Science and Engineering, University of Delaware, 201 DuPont Hall, Newark, Delaware 19716, United States.

*Correspondence should be addressed to: harrissb@ornl.gov or comes@udel.edu

[†]Authors contributed equally

**ABSTRACT**

Machine learning (ML) with *in situ* diagnostics offers a transformative approach to accelerate, understand, and control thin film synthesis by uncovering relationships between synthesis conditions and material properties. In this study, we demonstrate the application of deep learning to predict the stoichiometry of $Sr_xTi_{1-x}O_3$ thin films using reflection high-energy electron diffraction images acquired during pulsed laser deposition. A gated convolutional neural network trained for regression of the Sr atomic fraction achieved accurate predictions with a small dataset of 31 samples. Explainable AI techniques revealed a previously unknown correlation between diffraction streak features and cation stoichiometry in $Sr_xTi_{1-x}O_3$ thin films. Our results demonstrate how ML can be used to transform a ubiquitous *in situ* diagnostic tool, that is usually limited to qualitative assessments, into a quantitative surrogate measurement of continuously valued thin film properties. Such methods are critically needed to enable real-time control, autonomous workflows, and accelerate traditional synthesis approaches.

In recent years, machine learning (ML) has emerged as a promising tool in materials synthesis to discover or optimize the synthesis procedure for targeted structure, composition, or properties[1-3]. For synthesis, ML is typically used in an active-learning framework[4-7] to efficiently explore large synthesis parameter spaces and achieve the desired result in fewer experiments than traditional methods. In this approach, synthesis recipes are used to train ML models to predict one or more desired film characteristics, and Bayesian optimization is commonly used for sequential design of experiments. ML models can capture complex relationships in the data that correlate synthesis conditions and/or diagnostic measurements to material characteristics that may be otherwise difficult to discover or require significant domain expertise to reveal.

Synthesis of thin film materials using physical vapor deposition (PVD) methods such as molecular beam epitaxy[8] (MBE) and pulsed laser deposition[9] (PLD) are a cornerstone of discovery in materials science, allowing for the synthesis and study of a wide variety of solid-state materials and interfaces. These techniques possess large parameter spaces and are traditionally time-consuming and labor intensive, making their workflows prime candidates for modernization through application of ML and automation. Fully autonomous synthesis with magnetron sputtering[10] and PLD[11] have been realized by combing automated synthesis, characterization, and sample exchange equipment with ML but, so far, most PVD experiments are focused on using ML to streamline complex data analysis or assist in real-time control during growth of single samples using *in situ* diagnostics.

Historically, reflection high-energy electron diffraction[12] (RHEED) is a routine *in situ* diagnostic that can be used to examine the starting substrate and monitor the crystalline phase, orientation, and growth rate of subsequent thin films during MBE or PLD. In RHEED, a high energy (10-20 keV) beam of electrons is directed at the growing film's surface, where they interact with the top 2-5 nm of the material and scatter to form a diffraction pattern on a phosphor screen, which is typically recorded with a camera. Oscillations in the intensity or width of the specular reflection or diffraction spots is commonly used to monitor the growth rate during layer-by-layer growth[13]. The full RHEED pattern contains a wealth of additional information but is often limited to qualitative assessments due to the challenges of accurately simulating experimental patterns and the absence of comprehensive databases linking RHEED patterns to film properties beyond simple crystal structure or surface reconstructions. Such observations have been used to estimate $SrTiO_3$ stoichiometry in the past via MBE and PLD growth[14-16], but with a primary focus on whether films are Ti or Sr rich without quantifying the stoichiometry.

Most modern ML efforts with RHEED data have been for dimensionality reduction through matrix factorization techniques or pattern clustering[17-21], classification of the surface structure [22-25], sematic segmentation for diffraction pattern identification[26], and real-time changepoint identification[27]. Regression methods for linking RHEED images to continuous quantities - e.g. stoichiometry or conductivity - with deep learning remains a little explored area but is promising in terms of transforming RHEED into a makeshift surrogate measurement for more time-consuming or complicated physical property measurements. If proven effective, such models can be used in autonomous synthesis workflows or for real-time adjustment of deposition conditions. Notably, multivariate linear models using manually engineered RHEED features have been found to be effective predictors of stoichiometry in $W_{1-x}V_xSe_2$ films with a small amount of data[28]. Deep learning can circumvent the manual feature engineering step, which may fail to capture more subtle image features and relaxes the requirement for highly robust feature extraction (i.e. curve fitting), although its effectiveness has not yet been proven.

In this study, we seek to evaluate the efficacy of supervised deep learning with RHEED images for regression tasks. We trained a gated convolutional neural network (CNN) to predict the atomic fraction of Sr, $x$, from RHEED images of $Sr_xTi_{1-x}O_3$ thin films grown by PLD. We find that the model performs well on the regression task, predicting $x$ with reasonable accuracy with a dataset of only 31 samples. Further, we use explainable AI techniques with the trained model to aid us in revealing a previously unknown correlation between the intensity ratio and spacing of the (01) and (02) diffraction peaks with the cation stoichiometry in $SrTiO_3$. Thus, deep learning with RHEED images for regression tasks holds promise for prediction of other continuously valued film properties, rather than simple classifications, as well as an analysis tool to assist in interpreting RHEED pattern correlations to film properties. We anticipate that methods like this will become widely adopted as automated PLD and MBE equipment becomes more broadly available, which can produce a higher volume of reproducible samples for analysis that can reveal deeper insights to the synthesis process.

To evaluate the use of deep learning to predict the stoichiometry of thin films using RHEED images acquired during synthesis, we must generate our own data because no experimental database of well characterized materials with associated *in situ* diagnostics exists. We chose homoepitaxial $SrTiO_3$ as the test material system since its growth with PLD is well known and the cation stoichiometry is easily modulated via deposition from binary oxide targets. We synthesized 31 $Sr_xTi_{1-x}O_3$ films with various Sr/Ti ratios by sequentially depositing sub-monolayer thicknesses of SrO and $TiO_2$ from the respective targets. The procedure used here is similar to the one used by Herklotz et al.[29] RHEED was collected during deposition and the final images (after deposition was complete) were used as the inputs to the machine learning model.

After deposition, the atomic fraction of Sr was measured by XPS and was used as the model's prediction target.

Figure 1a shows the experimental arrangement where the SrO and TiO$_2$ targets are moved into the path of the ablation laser by rotating the target carousel. First, we determined the deposition rate of SrO and TiO$_2$ under the same deposition conditions to be 0.790 Å/pulse and 0.1865 Å/pulse corresponding to 3 and 11 pulses per half-unit cell of SrTiO$_3$, respectively. The deposition sequence for each film was multiple cycles of 3 SrO pulses and 11+N pulses of TiO$_2$ where N ranged from -5 to 10 to control the relative Sr flux fraction provided to the substrate per cycle. Figure 1b schematically shows the laser pulse sequence on each target for a single deposition cycle. Two films with different nominal thicknesses of 15 nm and 20 nm were deposited for each N value to check the repeatability of the measured stoichiometry and RHEED patterns.

Figure 1c is a typical example of the RHEED specular reflection intensity versus time along the STO [110] direction, taken from the N = -1 sample grown to 20 nm thick using 47 deposition cycles. Generally, the specular intensity drops immediately upon the first SrO pulse and, after the first few deposition cycles, either increases or decreases during deposition from each target with some evolution in the time during target exchange where no flux is incident on the substrate. The inset of Figure 1c highlights an intensity increase during SrO deposition (green shaded areas) followed by a further increase during the target exchange time which saturates to some value. The intensity decreases during TiO$_2$ deposition (blue shaded areas) with, again, some evolution in intensity during the target exchange time. There is a lower frequency beat with a period of ~6-8 cycles (in this example) that appears due to the SrO deposition either causing the intensity to decrease and then increase (green shaded area at ~79 s) or only increase (green shaded area at ~87 s) corresponding to SrO deposition onto different layer coverages[14,29] (i.e. incomplete half unit cell layers). Since each SrO/TiO2 deposition grows ~ 0.5 MLs of SrTiO3, we attribute the changes in intensity during deposition to the typical RHEED oscillations observed with a layer-by-layer growth mode where the minima and maxima correspond to ~0.5 MLs and a completed MLs respectively.

RHEED patterns along the [110] azimuth of SrTiO3 for the bare substrate, Sr rich (N = -5), stoichiometric (N = -1), and Ti rich (N=7) films are shown in Figure 1d-g. Past work has indicated that [110] azimuth is best for determination of stoichiometry in STO during shuttered MBE growth[14]. The films show a streaky pattern indicating layer-by-layer growth and faint Kikuchi lines in Figure 1e-f suggest good crystal quality in some films. The arrows highlight the streaks from the (01) and (02) diffraction planes. Finally, we compare the Sr flux fraction to the film stoichiometry, atomic fraction Sr, measured by XPS. The Sr flux fraction is given by the expression $N_{Sr} / (N_{Sr} + N_{Ti})$ where $N_{Sr}$ and $N_{Ti}$ are the number of atoms deposited per cycle determined from the deposition rates and number of pulses on each target. Sr atomic fraction is given by the same expression but using the peak areas of the Sr 3d and Ti 2p XPS core-levels. The Sr flux fraction is directly proportional to the atomic Sr fraction, with a slope of 1.01 ± 0.031 (including trivial endpoints of 0.0 and 1.0). Thus, we have a simple control parameter to vary the film stoichiometry, in this case, by selecting an N value to directly vary the amount of Ti supplied during growth of the $TiO_2$ sub-layer of $SrTiO_3$.

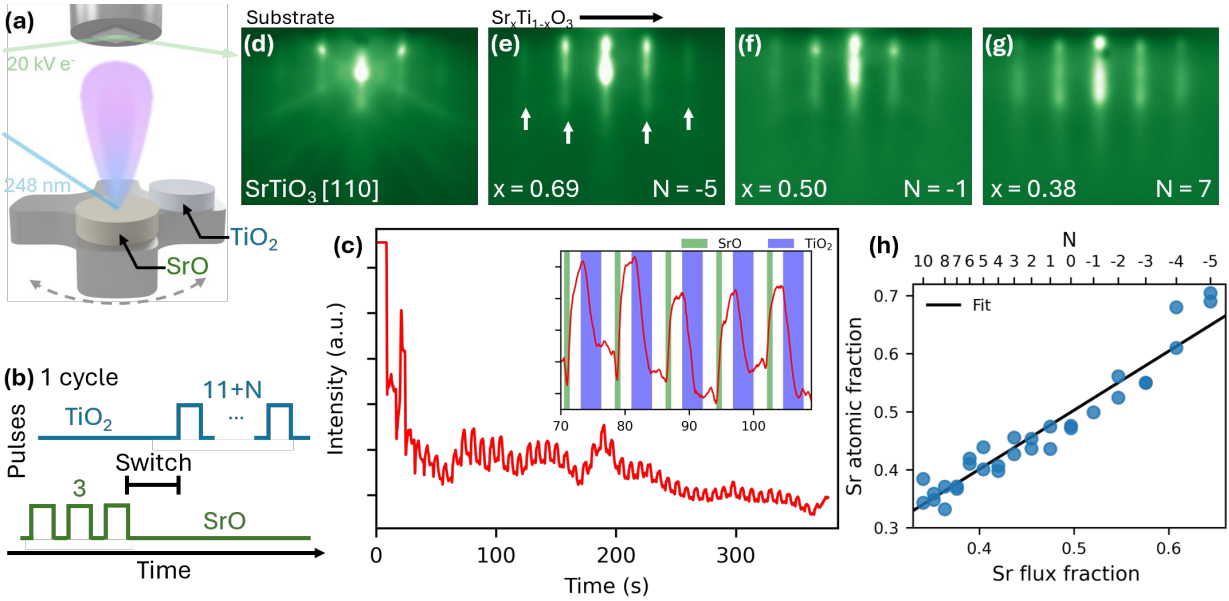

**Figure 1.** Deposition of $Sr_xTi_{1-x}O_3$ films by sequential pulsed laser deposition (PLD) of SrO and $TiO_2$ with reflection high-energy electron diffraction (RHEED) during growth. a) $Sr_xTi_{1-x}O_3$ films are grown by alternating sub-monolayer depositions from SrO and $TiO_2$ targets. b) Repeated cycles of 3 pulses of SrO and 11+N pulses of $TiO_2$ are used to grow films to nominal thicknesses of 15 nm and 20 nm where N ranges from -5 to 10. c) RHEED specular reflection intensity vs time for N = -1 shows oscillations modulated by the PLD target exchange time. The inset details the intensity modulation during deposition from each target. d) RHEED along the [110] azimuth of the $SrTiO_3$ substrate compared to the final RHEED image for e) Sr rich, f) stoichiometric, and g) Ti rich films along the same direction. The arrows indicate the (01) and (02) diffraction planes. h) The film Sr atomic fraction is directly proportional to the Sr flux fraction per deposition cycle with a fit slope of $1.01 \pm 0.031$.

Due to this direct correlation between the Sr flux and atomic fractions, the Sr flux fraction could be used as a "simple surrogate" machine learning target instead of the actual measured stoichiometry. This may be extensible to other materials synthesized through a similar binary target deposition method as an approximation for stoichiometry. However, for the most common single-target deposition scenario, it is not applicable. For example, the Sr/Ti ratio in films deposited from a pure STO target is known to change with the laser fluence and stoichiometry measurements would be required[30]. The stoichiometry resulting from various N values was highly repeatable, which is somewhat surprising considering the reproducibility problems that are known to exist in PLD and thin film growth in general. We ascribe the reproducibility, in this case, to our automated PLD system which carried out each deposition in a well-controlled, identical manner (at least in terms of the variables which were under our control). Additionally, the laser energy was measured *in vacuo* inside the chamber immediately before each deposition to remove the effect of laser window coating, which is not typically done in PLD due to technical challenges, and likely plays a role in the reproducibility considering the well-known stoichiometry modulation of STO with laser fluence.

Figure 2 shows characterizations of the selected films from Figure 1 to understand the surface quality, crystal structure, and chemistry of the $Sr_xTi_{1-x}O_3$ films used for machine learning. Figure 2a-c show the surface topography measured by atomic force microscopy (AFM) of Ti rich (N = 7), stoichiometric (N = -1), and Sr rich (N = -5) films. The films have a roughness of 0.547 nm, 0.492 nm, and 0.215 nm for N = 7, N = -1, and N = -5, respectively. This corroborates with the streaky RHEED patterns in Figure 1e-g and is consistent with a layer-by-layer growth process. X-ray diffraction (XRD) θ-2θ scans shown in Figure 2d confirm that the films are c-axis oriented $Sr_xTi_{1-x}O_3$ with no phase separation into the binary components $TiO_2$ or SrO due to sequential deposition from these targets. Figure 2e shows a detailed θ-2θ scan around the STO (002) reflection. We find an increase in the out-of-plane lattice parameter that is typical of off-stoichiometry STO films[31]. Finally, XPS spectra of the Sr 3d and Ti 2p core-levels (Figure 2f-g, respectively) confirm the presence of only $Sr^{2+}$ and $Ti^{4+}$ expected for STO. The binding energy for each core-level linearly increases with decreasing *x*. The Sr $3d_{5/2}$ peak has binding energy of 132.8 eV, 133.1 eV, and 133.4 eV for $x$ = 0.69, 0.50, and 0.38, respectively. Similarly, the Ti $2p_{3/2}$ peak has a binding energy of 456.9 eV, 457.2 eV, and 457.6 eV for the same $x$ values, respectively. These shifts are most likely due to variation in charge compensation that is not fully corrected by aligning the C 1s peak to 285.0 eV and do not indicate changes in Sr or Ti ionic charge state. Note here that the fit areas of these XPS peaks are the ones used for each sample to calculate the Sr atomic fraction used as the machine learning target. It is worth noting that the value of $x$ extracted from XPS fitting is not necessarily a precise estimate of film composition due to surface effects, such as the preferential accumulation of Sr or Ti on the surface of Sr-rich or Ti-rich films, respectively[32]. However, the relationship between absolute composition, such as what might be measured via Rutherford backscattering, and XPS-derived composition should be monotonic and does not affect the validity of the ML model.

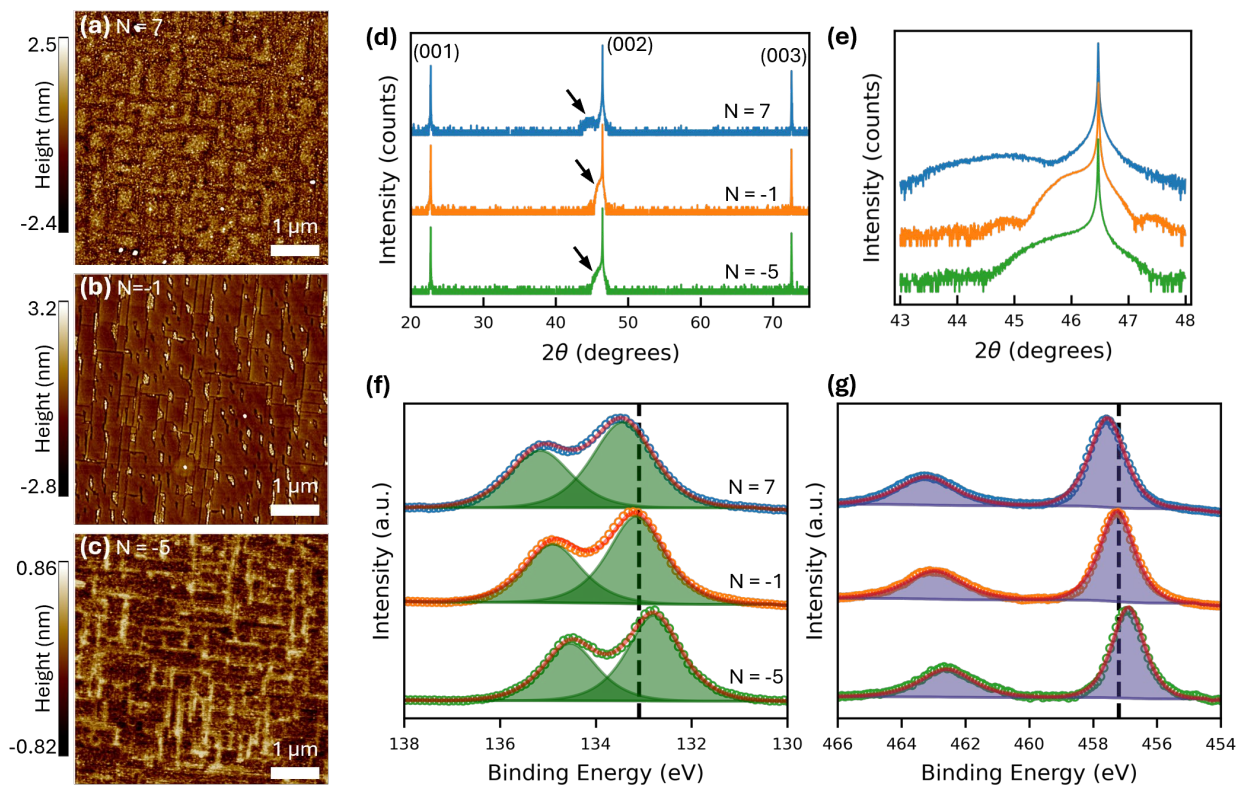

**Figure 2.** Characterization of selected $Sr_xTi_{1-x}O_3$ films that are Sr rich (N = -5), stoichiometric (N = -1), and Ti rich (N = 7). a-c) Film topography from atomic force microscopy images display atomically flat surfaces. d) Wide angle θ-2θ XRD scans confirm (001) oriented $Sr_xTi_{1-x}O_3$ films with no impurity $TiO_2$ or SrO phases present. e) θ-2θ XRD scan of the $SrTiO_3$ (002) reflection shows c-axis expansion the film when for Sr-rich and Ti-rich cases. XPS of the f) Sr 3d and g) Ti 2p core-levels show examples of the calculated peak areas used to measure the atomic fraction of Sr in each film. XPS also shows a linear shift in the binding energies of Sr 3d and Ti 2p with stoichiometry. The dashed lines mark the typical binding energies of 133.1 eV and 457.2 eV for the Sr $3d_{5/2}$ and Ti $2p_{3/2}$ peaks in $SrTiO_3$, respectively.

Using the dataset we generated for $Sr_xTi_{1-x}O_3$ films, we trained a gated convolutional neural network (CNN) to extract features from the RHEED images to predict the Sr atomic fraction in the films. Instead of using standard convolutions, we chose gated convolutions[33,34] to leverage the effective spatial attention mechanism created by the learned gate weights. Figure 3a shows the model architecture which consisted of 3 sequential blocks of gated convolutions and max pooling which is finally flattened and passed through 2 gated linear units with dropout to output the Sr atomic fraction prediction. The gated convolution splits the output of a standard 2D convolution along the channel dimension, applies a nonlinear activation function to one half, a sigmoid function to the other (the gate), and multiplies the results to rescale the convolution activations based on the learned spatial attention of the gate. The model hyperparameters were optimized through a 3-fold cross-validation tuning procedure, with both image and target augmentation, to identify the optimal configuration. Figure 3b shows the L1 loss curve of the optimized model, demonstrating its convergence. Figure 3c shows the model predicted vs actual Sr atomic fraction for the training and validation sets. The model achieves a coefficient of determination of $r^2_{train} = 0.956$ and $r^2_{val} = 0.867$, respectively, demonstrating reasonable accuracy on this small dataset of 31 samples. Thus, the model can effectively learn deep features from the RHEED images that are highly correlated to the film stoichiometry.

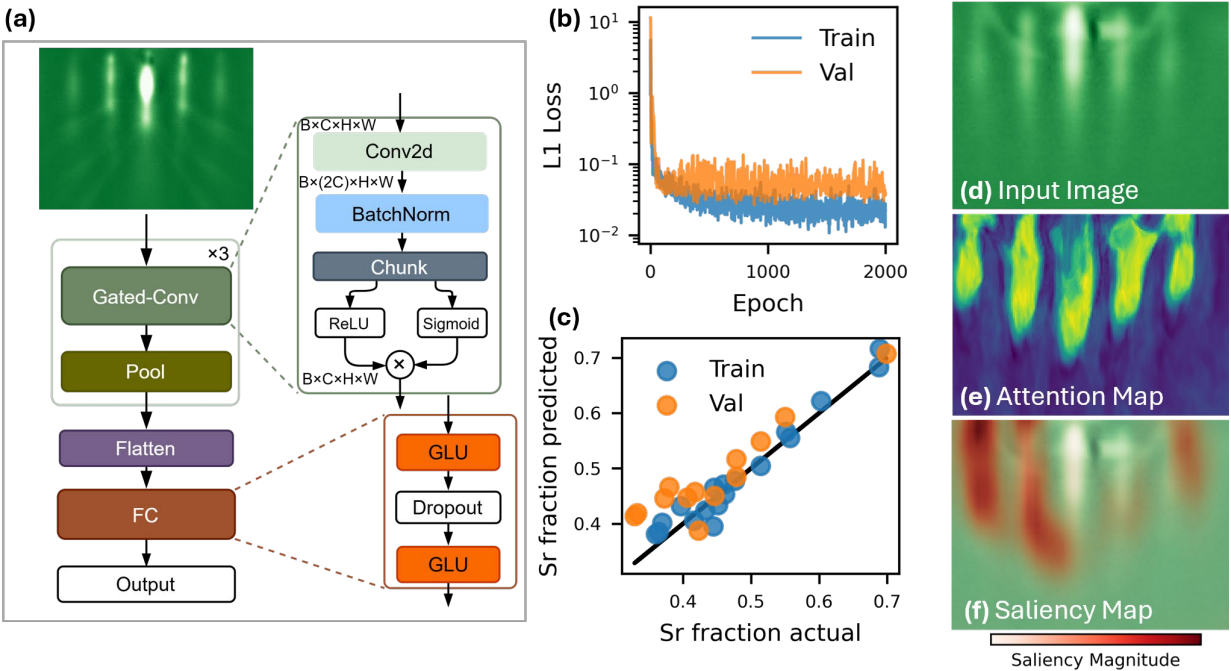

**Figure 3.** Deep learning of RHEED images for prediction of Sr atomic fraction in $Sr_xTi_{1-x}O_3$ films. a) The machine learning model extracts deep features by passing RHEED images through 3 sequential gated convolutional layers, where the gate in each layer acts as an attention mechanism that emphasizes the most important feature activations. b) Learning curve of L1 loss vs. training epoch shows convergence. c) Model predicted vs. actual Sr atomic fraction with coefficient of determination $r^2_{train} = 0.956$ and $r^2_{val} = 0.867$. For an example input image (shown in d), the e) attention map derived from the mean gate weights highlights the important regions in the image and the f) saliency map indicates the pixel importance of the $Sr_xTi_{1-x}O_3$ (01) and (02) RHEED streaks for predicting the Sr atomic fraction.

To understand what regions of the RHEED images the model correlates to the Sr atomic fraction, we visualize attention maps and employ gradient-weighted class activation mapping, Grad-CAM++[35] modified for regression, to visualize the pixel saliency. Using an example image (Figure 3d), we visualize the attention map by averaging all 128 gate channels of the final convolution layer, shown in Figure 3e. The attention mechanism primarily gives large weight the diffraction streaks and small weight to the background in the lower portion of the image and between the streaks. Interestingly, the attention mechanism gives no weight to an image artifact created by a defect in the RHEED screen (seen as the dark spot in Figure 3d), which is consistent because it has no bearing on the film properties. Finally, we visualize the pixel saliency, which highlights the pixels that are most influential to the prediction. A large saliency magnitude indicates that the pixel value (intensity) at a given location greatly affects the prediction. Figure 3f shows the saliency map, with most of the saliency weighted towards the (01) and (02) diffraction planes, giving little weight to the central streak or spectral reflection. Generally, the model predictions are influenced by the diffraction streaks and the greatest salience typically involves the pixels containing or near the (01) and (02) diffraction planes.

Finally, we can use the information learned from the saliency maps to guide manual analysis of the RHEED images to search for empirical correlations to corroborate/explain the model predictions. We integrated the RHEED images and fit Gaussians to the diffraction peaks to examine the relative intensity and spacing (i.e. in-plane lattice parameter) of the (01) and (02) planes. Figure 4a shows an example of the integrated RHEED with the multi-Gaussian fit. Indeed, we find empirical correlations between the (01) and (02) peak parameters and the Sr atomic fraction. Figure 4b shows the (01)/(02) intensity ratio vs Sr fraction. We clearly observe a strong positive correlation, although with an outlier point. Similarly, Figure 4c shows the ratio of the pixel spacing of each peak relative to the central diffraction spot, i.e. $(C_{(02)} - C_{(11)})/(C_{(01)} - C_{(11)})$ where C is the peak center. This is equivalent to taking the ratio of the in-plane lattice parameter as calculated from the 1st and 2nd order diffraction planes. We also observe a positive correlation with this metric. To our knowledge, these empirical correlations between diffraction intensity/spacing ratios and Sr/Ti ratio in STO have not been noted in the literature, likely due to a lack of data which spans such a broad range of stoichiometry. We speculate that the cause of these variations is due to distortions from the ideal $SrTiO_3$ lattice, which alter the scattering intensity of diffraction planes.

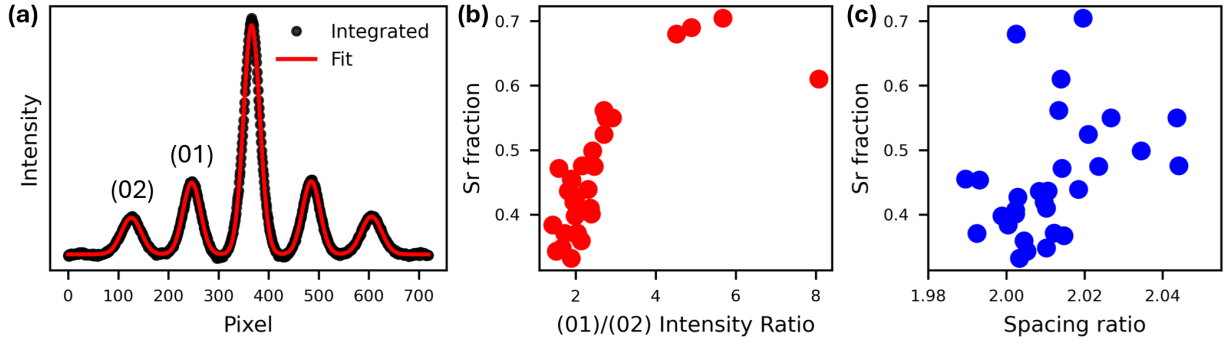

**Figure 4.** Empirical RHEED features suggested from salience analysis of machine learning model that are highly correlated to the atomic fraction of Sr in $Sr_xTi_{1-x}O_3$ thin films. a) Example of an integrated RHEED image for one sample (N = 6, $x$ = 0.419) with a multiple Gaussian peak fit. b) The intensity ratio of the (01)/(02) RHEED diffraction streaks and the c) The ratio of the pixel spacing from the central diffraction plane of the (02) and (01) streaks. Both empirical metrics were discovered with the aid of the deep features learned from the model.

In this work, we demonstrated the efficacy of deep learning for regression of continuous valued properties of thin films using RHEED images acquired during synthesis. We synthesized 31 $Sr_xTi_{1-}$ $_x$O$_3$ thin films using PLD, recorded RHEED images for each sample during deposition, and measured the atomic fraction of Sr using XPS to generate a dataset for analysis. Using this dataset, we trained a gated convolutional neural network that can successfully predict the Sr atomic fraction from the RHEED images, with an accuracy that is comparable to post-growth XPS measurements. Explainable AI techniques indicated a correlation between the (01) and (02) diffraction planes and the film stoichiometry. Guided by these machine-learned features, manual analysis revealed previously unknown empirical correlations between the relative intensity and spacing of these diffraction peaks. Our findings highlight the potential for deep learning with RHEED images as a "simple surrogate" measurement that can substitute for more in-depth *ex situ* properties measurements that are affected by crystal structure. Such an approach can accelerate both traditional and autonomous synthesis campaigns by deriving greater value from otherwise qualitative routine diagnostics where correlations are not obvious or previously known.

**CONFLICT OF INTEREST**

The authors declare no competing interests.

**AUTHOR CONTRIBUTIONS**

**S.B.H.**: investigation (lead); data curation (lead); methodology (equal); software (lead); visualization (lead); writing—original draft (lead); writing—review & editing (lead). Conceptualization (supporting). **P.T.G.**: Conceptualization (equal); investigation (supporting); software (supporting); writing—original draft (supporting); writing—review & editing (supporting). **C.M.R.**: Resources (equal); Methodology (equal); writing—review & editing (supporting). **R.K.V.**: writing—review & editing (supporting); Conceptualization (equal); Investigation (supporting) **R.B.C.**: Conceptualization (equal); writing—review & editing (supporting), Resources (equal). All authors read and approved the final manuscript.

## DATA AVAILABILITY

The data that support the findings of this study are openly available at https://github.com/sumner-harris/Deep-Learning-with-RHEED.git

## CODE AVAILABILITY

The data that support the findings of this study are openly available at https://github.com/sumner-harris/Deep-Learning-with-RHEED.git

## ACKNOWLEDGEMENTS

This work was supported by the Center for Nanophase Materials Sciences (CNMS), which is a US Department of Energy, Office of Science User Facility at Oak Ridge National Laboratory. P.T.G. and R.B.C. gratefully acknowledge funding for RHEED analytics from the National Science Foundation Division of Materials Research under award DMR-2045993. P.T.G. also acknowledges support from the Department of Energy's Office of Science Graduate Student Research Program (DE-SC0014664).

**Supporting Information Available**

Experimental methods; Machine learning model training and data preprocessing.

# Supplemental Information for:

# Deep learning with reflection high-energy electron diffraction images to predict cation ratio in $Sr_{2x}Ti_{2(1-x)}O_3$ thin films

*Sumner B. Harris[1*†], Patrick T. Gemperline[2†], Christopher M. Rouleau[1], Rama K. Vasudevan[1], Ryan B. Comes[2,3*]*

4. Center for Nanophase Materials Sciences, Oak Ridge National Laboratory, Oak Ridge, Tennessee 37831, United States.
5. Department of Physics, Auburn University, 315 Roosevelt Concourse, Auburn, Alabama 36849, United States.
6. Department of Materials Science and Engineering, University of Delaware, 201 DuPont Hall, Newark, Delaware 19716, United States.

*Correspondence should be addressed to: harrissb@ornl.gov or comes@udel.edu

†Authors contributed equally

## METHODS

### Experimental

PLD of $Sr_xTi_{1-x}O_3$ films was performed by sequentially depositing sub-monolayer (MLs) thick layers of SrO and $TiO_2$ through ablation of the respective targets with a KrF excimer laser (Coherent LPX 305F, 248 nm, 25 ns). A 10×10 mm aperture was imaged by a projection beamline onto the targets to produce a rectangular laser spot with an area of 0.0351 cm$^2$. The laser energy was fixed at 35 mJ/pulse, resulting in a fluence of 1.0 J/cm$^2$. A repetition rate of 3 Hz was used, the target-substrate distance was 4.5 cm, and the substrate temperature was 700 °C. The base pressure for each deposition was < 3×10$^{-6}$ Torr. The substrates were heated and the films were deposited in 20 mTorr $O_2$ (99.9999%, 5 sccm). The 5×5 mm (001) $SrTiO_3$ substrates (CrysTec GmbH) were prepared by annealing at 1050 °C in air followed by deionized water leaching[1] for 30 s and a final anneal at 1050 °C to produce a stepped, $TiO_2$ terminated surface. RHEED patterns were generated with an electron beam (Staib TorrRHEED, 20 kV, 1.4 mA emission current) and phosphor screen, and captured at 10 fps (5 ms exposure time) during deposition with a 16-bit CMOS camera (FLIR Blackfly S BFS-PGE-04S2M-CS, 291 frames per second (fps), 720×540 pixels).

XPS was performed at 2×10$^{-10}$ Torr using a monochromated Al Kα x-ray source (1486 eV) with a take-off angle normal to the sample surface. The SPECS Phoibos 150 hemispherical analyzer was operated at 40 eV pass energy for high resolution scans. An electron flood gun was used for charge neutralization and the C 1s line was referenced to 285.0 eV for binding energy correction. The elemental composition was estimated using the peak areas of the Sr 3d and Ti 2p core levels. The atomic fraction of Sr was calculated by $x = A_{Sr} / (A_{Sr} + A_{Ti})$ where $A_{Sr}$ and $A_{Ti}$ are the areas of the Sr 3d and Ti 2p peaks, respectively. High resolution x-ray diffraction (XRD) was performed using a PANalytical X'pert PRO MRD diffractometer using a hybrid monochromator (Cu Kα1) and a Ge (220) analyzer crystal. Atomic force microscopy (AFM) was performed with a Bruker Dimension Icon in tapping mode with a Si probe (TESPA-V2, 7 nm tip radius, 37 N m$^{-1}$ spring constant.)

**Machine learning model training and data preprocessing**

Each RHEED image has 1 channel (16-bit) with 720×540 pixels and is associated with 1 value of Sr atomic fraction. Each image has a broad background from the substrate heater (diffuse infrared light scattering) which is removed by subtracting a baseline calculated with the asymmetric least squares algorithm[2] from each row of the image. Alternatively, a background can be created via gaussian blur with a large variance and subtracted from the image and is more computationally efficient for the case of on-line image processing. For training, the images were downsized by a factor of 4 to 135×180 pixels, the log of the intensity was taken to enhance weak features, and the intensity was min-max normalized. We used a simple machine learning model, composed of 3 gated 2D convolution layers, each followed by a max pooling layer with a (2×2) kernel and the output of layer 3 used an adaptive max pooling layer to reduce the output to be the same as the number of channels. This output was flattened and passed through 2 gated linear unit layers with a dropout layer between and the final number of outputs was set to 1 for regression.

The data was shuffled and split 70/30 into training and validation sets with 21 and 10 images, respectively, and training is done with full batch gradient decent. To increase the variety in this small dataset for improved model generalization, we augment both the images and the stoichiometry targets in each training epoch. The images are randomly transformed with vertical and horizontal flips, ± 5° degree rotation, and rescaled between 95-130%. These transformations are meant to mimic a variety of experimental RHEED geometries and simulate small lateral sample tilt and camera zoom. The Sr atomic fraction is augmented by adding normally distributed noise with 0 mean and 0.015 variance, which simulates the typical ± 2.5 atomic percent error in XPS measurements. We conducted extensive hyperparameter tuning using 3-fold cross validation with L1 loss (for outlier resilience) using Ray Tune[3] with the Optuna algorithm[4]. The hyperparameters that were explored were the Adam[5] learning rate, number of channels in the first convolution layer (start channels), convolution kernel size, dropout rate, and pooling layer type (either max or average pooling). The optimal model had 16 start channels (multiplied by 4 and 8 for layers 2 and 3 respectively), kernel size of (7×7), dropout rate of 0.152, max pooling layers, and a learning rate of 0.0249. The optimized model scores a mean coefficient of determination $r^2 = 0.83$ on the validation set, which is the mean of validation $r^2$ for each of the 3 folds. This can be considered a small model, containing only 906k parameters (3.64 MB), which can be easily run on modern personal computers or low-cost single board machines.